# Vector spectrometer with Hertz-level resolution and super-recognition capability

Ting Qing[1,2,6], Shupeng Li[1,5,6], Huashan Yang[1], Lihan Wang[1], Yijie Fang[1], Xiaohu Tang[1], Meihui Cao[1], Jianming Lu[2], Jijun He[1], Junqiu Liu[4], Yueguang Lyu[3,*], Shilong Pan[1,*]

[1]National Key Laboratory of Microwave Photonics, Nanjing University of Aeronautics and Astronautics, Nanjing 211106, China
[2]2012 Lab, Huawei Technologies Co., Ltd, Shenzhen, China
[3]School of Physics, Harbin Institute of Technology, Harbin, China
[4]International Quantum Academy, Shenzhen 518048, China
[5]Newkey Photonics Technologies Co., Ltd., Suzhou 215500, China
[6]These authors contributed equally: Ting Qing, Shupeng Li
*Corresponding author. Email: pans@nuaa.edu.cn, boblu@vip.163.com

**Abstract**
High-resolution optical spectrometers are crucial in revealing intricate characteristics of signals[1], determining laser frequencies[2], measuring physical constants[3], identifying substances[4], and advancing biosensing applications[5]. Conventional spectrometers, however, often grapple with inherent trade-offs among spectral resolution, wavelength range, and accuracy. Furthermore, even at high resolution, resolving overlapping spectral lines during spectroscopic analyses remains a huge challenge. Here, we propose a vector spectrometer with ultrahigh resolution, combining broadband optical frequency hopping, ultrafine microwave-photonic scanning, and vector detection. A programmable frequency-hopping laser was developed, facilitating a sub-Hz linewidth and Hz-level frequency stability, an improvement of four and six orders of magnitude, respectively, compared to those of state-of-the-art tunable lasers. We also designed an asymmetric optical transmitter and receiver to eliminate measurement errors arising from modulation nonlinearity and multi-channel crosstalk[6]. The resultant vector spectrometer exhibits an unprecedented frequency resolution of 2 Hz, surpassing the state-of-the-art by four orders of magnitude, over a 33-nm range. Through high-resolution vector analysis, we observed that group delay information enhances the separation capability of overlapping spectral lines by over 47%, significantly streamlining the real-time identification of diverse substances. Our technique fills the gap in optical spectrometers with resolutions below 10 kHz and enables vector measurement to embrace revolution in functionality.

**Main**
Over the past century, optical spectrometers have undergone remarkable advancements,

emerging as non-invasive and efficient measurement tools across diverse fields, encompassing physics, chemistry, material science, environmental technology, astronomy, biology, and medicine. However, the limited resolution of spectrometers poses a bottleneck to the further progression of applications in these domains. For instance, by capturing minute resonance shifts or mode splitting of optical microcavities induced by nanoscale objects[5], spectrometers can be used for nanoparticle analysis and biomolecular detection. These optical microcavities pursue a Q factor up to $3\times10^{11}$ at 1550 nm (spectral width<1 kHz)[7] to improve the detection sensitivity, requiring a sub-kHz measurement resolution. Similarly, narrowband phenomena, such as a 36-Hz-width spectral hole produced by the quantum coherence effect in a ruby crystal[8], and narrow-linewidth signals, such as a 5-Hz-linewidth optical frequency comb[9] and Hz-linewidth lasers[10], have the capability of fine spectral manipulation. These narrowband devices can improve the performance of coherent optical communications[11], optical sensing[12], optical atomic clocks[13], light detection and ranging (LIDAR) [14], and photonic microwave generation[15], but require spectrometers with Hz-level resolution. The high resolution also helps to determine the crucial physical constants[16], such as the Boltzmann constant[3] and the Rydberg constant[16].

The pursuit of enhanced resolution in spectral measurement has been a longstanding research challenge, evolving from the realm of a few GHz with diffraction grating spectrometers[17-19] and Fourier spectrometers[20-21], progressing to 100 MHz with Fabry-Pérot spectrometers[22-23], and further to 5 MHz representing the best resolution of commercial spectrometers based on swept-frequency interferometry (SFI)[24]. Even finer resolutions, down to 30 kHz[1], have been achieved with spectrometers based on stimulated Brillouin scattering (SBS)[1,25]. However, the relentless quest for superior resolution has encountered significant bottlenecks. To achieve Hz-level resolution, spectrometers based on gratings, Fabry-Pérot interferometers, and SBS necessitate tunable optical filters with bandwidths at the Hz level, while Fourier spectrometers demand adjustable delay on the order of seconds (equivalent to 300 000 km in length), which verges on impracticability. SFI-based spectrometers provide an alternative, with resolution depending on the linewidth of tunable lasers (TLSs), frequency tuning precision, and frequency stability. However, to achieve Hertz-level resolution in a wide frequency range, several challenges must be overcome: improving the fineness of frequency tuning, compressing the laser linewidth, stabilizing the laser frequency, and breaking the mutual constraint between tunability and stability. These challenges have, until now, restricted spectrometers to either relatively low resolution over a wide frequency range or a high resolution within a narrow frequency range.

High-resolution optical vector analysis (OVA)[6,26-34] based on microwave photonics (MWP) offers a potential solution to overcome the above challenges. Originally proposed for measuring the spectral response of narrowband passive devices, MWP-based OVA utilizes electro-optical (EO) modulation to map hyperfine frequency scanning from the electrical domain to the optical domain. This approach, combined with spectral channelization leveraging optical frequency comb (OFC)[6], allows for achieving high resolution and wide range simultaneously. Notably, OVA demonstrates the capability for high-resolution vector measurement, capturing parameters like magnitude, phase, and group delay. However, MWP-based OVA faces limitations in measuring the frequency and phase of active devices (such as laser sources) due to the use of direct detection. Moreover, nonlinearities in the EO modulation, along with channel crosstalk stemming from OFC unevenness and inadequate out-of-band suppression of tunable optical filters used for mode selection, would introduce considerable measurement errors[6]. These issues significantly prevent existing MWP-based OVA from demonstrating resolution superiority over traditional spectrometers in practical spectral measurement.

Here, we propose a vector spectrometer for high-resolution spectral analysis of both active and passive devices. The spectrometer seamlessly integrates broadband optical frequency hopping with ultrafine microwave photonic frequency scanning, enabling it to achieve a verified frequency resolution of 2 Hz and a measurable spectral range of 33 nm. This integration effectively overcomes the historical compromise between frequency resolution and measurement range. The key to achieving the 2-Hz resolution lies in the implementation of a frequency-hopping laser with a sub-Hz linewidth and Hz-level frequency stability. To realize the laser, we first lock the central comb line and mode spacing of a microcomb to a sub-Hz linewidth ultrastable laser and an atomic clock, respectively. Then, we offset-lock a TLS to the stabilized microcomb and program it to achieve flexible frequency hopping. This innovation breaks the mutual constraint between tunability and stability. The adoption of the frequency-hopping laser replaces traditional OFC-based spectral channelization, thereby eliminating channel crosstalk. To further remove the measurement errors induced by EO modulation nonlinearity, the vector spectrometer employs a parallel asymmetric spectrum measurement architecture on the basis of our previous OVA[6]. The architecture consists of a pair of asymmetric optical double-sideband modulation signal generator and receiver. When measuring an active device, the receiver switches to coherent detection mode. Combined with the local reference signal generated from the stable sub-Hz linewidth frequency-hopping laser, the measurement system accurately extracts the amplitude and phase of the optical signal from the active device. Furthermore, the vector spectrometer can accurately measure the group delay response. We found that the full width at half maximum (FWHM) of this response is at least 47% narrower than the magnitude response, which holds immense potential for the real-time separation of overlapping absorption lines in multi-component systems. We foresee that this innovative spectrometer design will pave the way for high-performance solutions in next-generation spectral analysis.

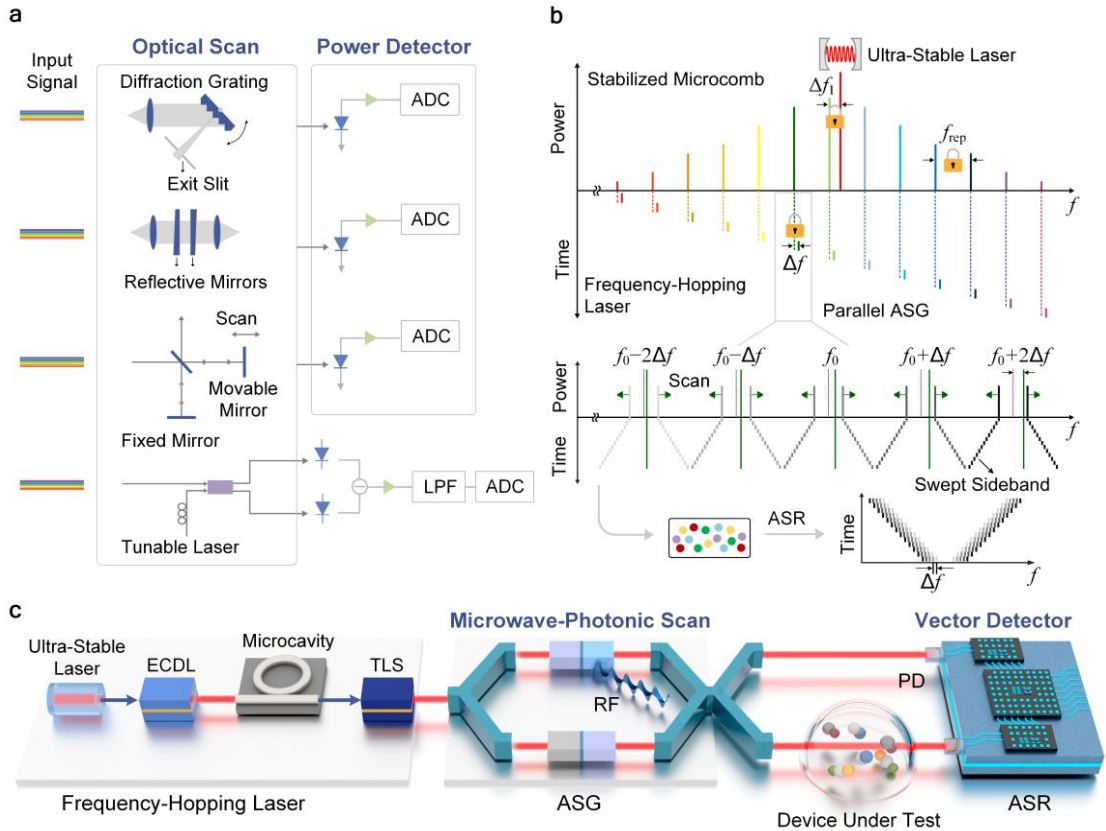

**Fig. 1 Concept of the vector spectrometer and its operational principle. a,** Schematic of four common types of spectrometers. These spectrometers typically utilize optical scanning and power detectors, inherently limiting their wavelength resolution and the ability to measure vectorial information. **b,** Illustration of the principle of the vector spectrometer. The central line and comb spacing of a soliton microcavity comb are locked to an ultra-stable laser with a sub-Hz linewidth and an atomic clock, respectively. A TLS is offset-locked to the microcomb and is programmed to provide the flexibly-hopping frequency. The parallel ASG generates two EO combs, with five comb lines in the middle having frequency differences of $f_0-2\Delta f$, $f_0-\Delta f$, $f_0$, $f_0+\Delta f$, and $f_0+2\Delta f$, respectively. One EO comb is modulated by a microwave signal via carrier-suppressed modulation, generating ten sidebands for channelized measurements. Microwave stepped frequency scanning results in parallel scanning of the ten channels, enabling broadband coverage with ultrahigh resolution. After photodetection, RF spikes with stepped sweeping frequencies are obtained, allowing extraction of the spectral information of the device under test covered by the ten sidebands (approximately 103 GHz). **c**, Schematic of the vector spectrometer using the frequency-hopping laser, microwave-photonic scan, and vector detector. An ECDL pumps an integrated $Si_3N_4$ microcavity, generating a soliton comb. An ultra-stable laser serves as an optical reference for the microcomb. By employing an offset phase-locked loop to synchronize a TLS with one comb line, a fast-switching, stable, and ultra-narrow-linewidth frequency-hopping laser is obtained. The frequency-hopping laser is connected to a parallel ASG to generate probe signals. After device interrogation, a parallel ASR extracts the desired spectral information from the probe signals. ADC,

analog to digital converter; LPF, low-pass filter; ECDL, external cavity diode laser; TLS, tunable laser source; ASG, asymmetric optical signal generator; ASR, asymmetric optical signal receiver; RF, radio frequency; PD, photodetector.

**Principle of the vector spectrometer**

Figure 1 illustrates the concept of the vector spectrometer and its operational principle. In Fig. 1a, common types of spectrometers using optical scanning and power detectors are depicted. Optical scanning offers broad coverage but limited resolution, while power detectors can only perform scalar measurements for magnitude. Fig. 1b presents the principle illustration of the vector spectrometer (see Fig. 1c). To implement the frequency-hopping laser, first, we generate a soliton microcomb[35-39] by pumping a $Si_3N_4$ microresonator using a continuous-wave light. The central line and comb spacing of the microcomb are locked to an ultra-stable laser with a sub-Hz linewidth and an atomic clock via two offset phase-locked loops, respectively, resulting in a fully stabilized soliton microcomb—a precise spectral ruler. Then, by using another offset phase-locked loop, we lock the frequency difference between a TLS and one comb line of the fully stabilized microcomb to a microwave reference. This frequency difference, determined by the microwave reference, can be finely tuned in steps of 1 Hz. If the frequency of the microwave reference can be tuned in a range exceeding half of the microcomb repetition, the locked TLS can generate any desired optical frequencies with a setting resolution of 1 Hz within the frequency range covered by the microcomb. Compared with conventional TLS sources, the frequency-hopping laser offers two unparalleled advantages. Firstly, it provides more precise frequency accuracy across a broad wavelength range. Secondly, it has a much faster frequency-switching speed. Since the frequency-hopping laser only needs to stabilize the microcomb once, it can swiftly lock to the set frequency (typically sub-second), whereas a TLS source needs to stabilize the frequency every time it switches (typically tens of seconds).

Then, the frequency-hopping laser is directed to a parallel ASG to generate asymmetric optical probe signals. The fundamental concept of the parallel ASG is to produce two EO combs with slightly different central frequencies (differing by $f_0$) and repetition frequencies (varying by $\Delta f$). One EO comb generates five carrier-suppressed optical double-sideband (CS-ODSB) modulated signals, while the other serves as five frequency-shifted carriers. By sweeping the frequency $f_e$ of the modulation signal, the probe signals, i.e., the ten sidebands, interrogate the device under test and retrieve their spectral information within a frequency span equivalent to half of the EO comb spacing. After square-law detection in a photodetector, RF spikes are produced at the frequencies of $f_e-f_0-(m-1)\cdot\Delta f$ and $f_e+f_0+(m-1)\cdot\Delta f$ ($1\leq m\leq N$, where $N$ is the number of comb lines in a comb). By judiciously selecting $f_0$ and $\Delta f$, all these components can be effectively separated in the electrical spectrum. The parallel ASR utilizes electrically tunable bandpass filters to select the desired frequency components and extract the spectral information carried by the probe signals, eliminating the influence of high-order sidebands and other unwanted components. By sweeping the output frequency of the frequency-hopping laser and subsequently measuring the device under test by the ASG and ASR, the measurement range of the proposed vector spectrometer can reach the frequency range covered by the microcomb.

**The frequency-hopping laser**

Here, we present a novel approach for establishing a frequency-hopping laser by precisely locking a tunable laser to the comb lines of a fully stabilized dissipative Kerr soliton (DKS) comb. The experimental arrangement is depicted in Fig. 2a. A $Si_3N_4$ microresonator is pumped with an external cavity diode laser (ECDL) that has been amplified by an erbium-doped fiber amplifier (EDFA). By scanning the ECDL (Toptica CTL 1550) across the resonance, we generate a soliton microcomb. As illustrated in Fig. 2b, this soliton microcomb has a $sech^2$-like envelope with a mode spacing of approximately 103 GHz. To ensure comprehensive stabilization of the microcomb, we employ two offset phase-locked loops, meticulously locking the central comb line and the repetition frequency to an ultra-stable laser and an atomic clock, respectively. This synchronization is achieved by actively controlling the frequency and power of the pump laser. The ultra-stable laser (MenloSystems ORS) is distinguished by its sub-Hz linewidth and stability of $1.26\times10^{-15}$ at a 1-second integration time. To monitor the frequency drift of the central comb line for feedback control, we select the comb line using a fiber Bragg grating (FBG). The selected comb line is then made to beat with the ultra-stable laser. The obtained beat note is mixed with a reference RF signal, the frequency of which is fixed at $\Omega$. After low-pass filtering, the down-converted signal remains constant only when the frequency difference between the central comb line and the ultra-stable laser equals $\Omega$. Therefore, this down-converted signal can serve as an error signal for precise phase stabilization. A proportional-integral-derivative controller is then employed, which outputs a control signal to the frequency modulation input port of the ECDL to lock the central comb line to the ultra-stable laser.

To quantitatively assess the stability of the central comb line, we measured the out-of-loop beat note between the central comb line and the ultra-stable laser. As depicted in Fig. 2c, the spectrogram of the free-drifting beat reveals an approximately linear frequency drift, reaching approximately 6 MHz within one minute. In stark contrast, the stabilized beat displays a frequency drift of less than ±0.7 Hz over a span of 1000 s, as evident in Fig. 2d. Furthermore, the stabilized beat showcases a linewidth of roughly 1 Hz, as illustrated in Fig. 2e. These results underscore the remarkable stability and precision achieved through our frequency-hopping laser, offering significant implications for high-resolution spectroscopy and precision frequency control.

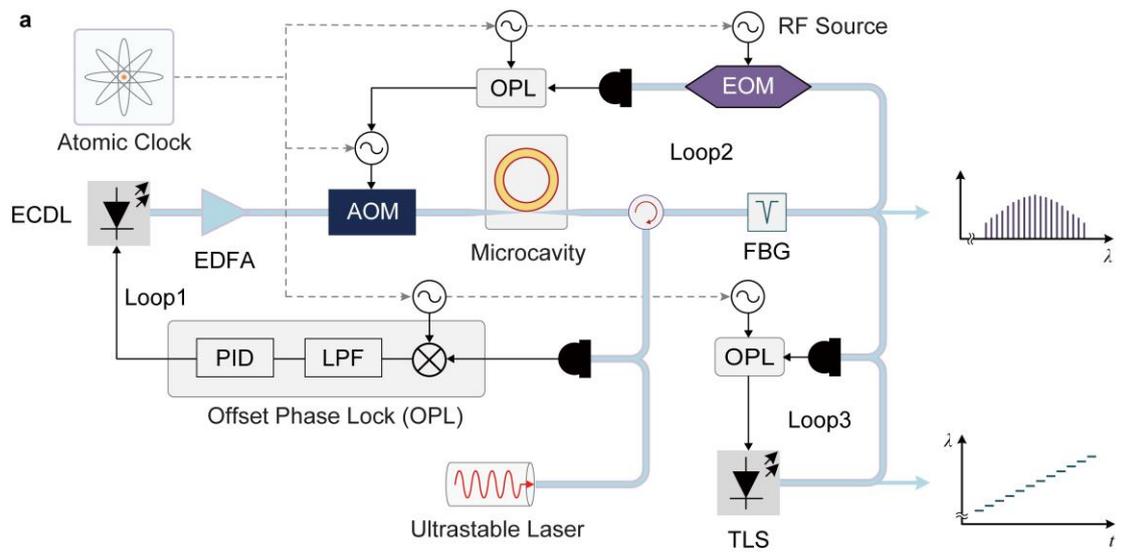

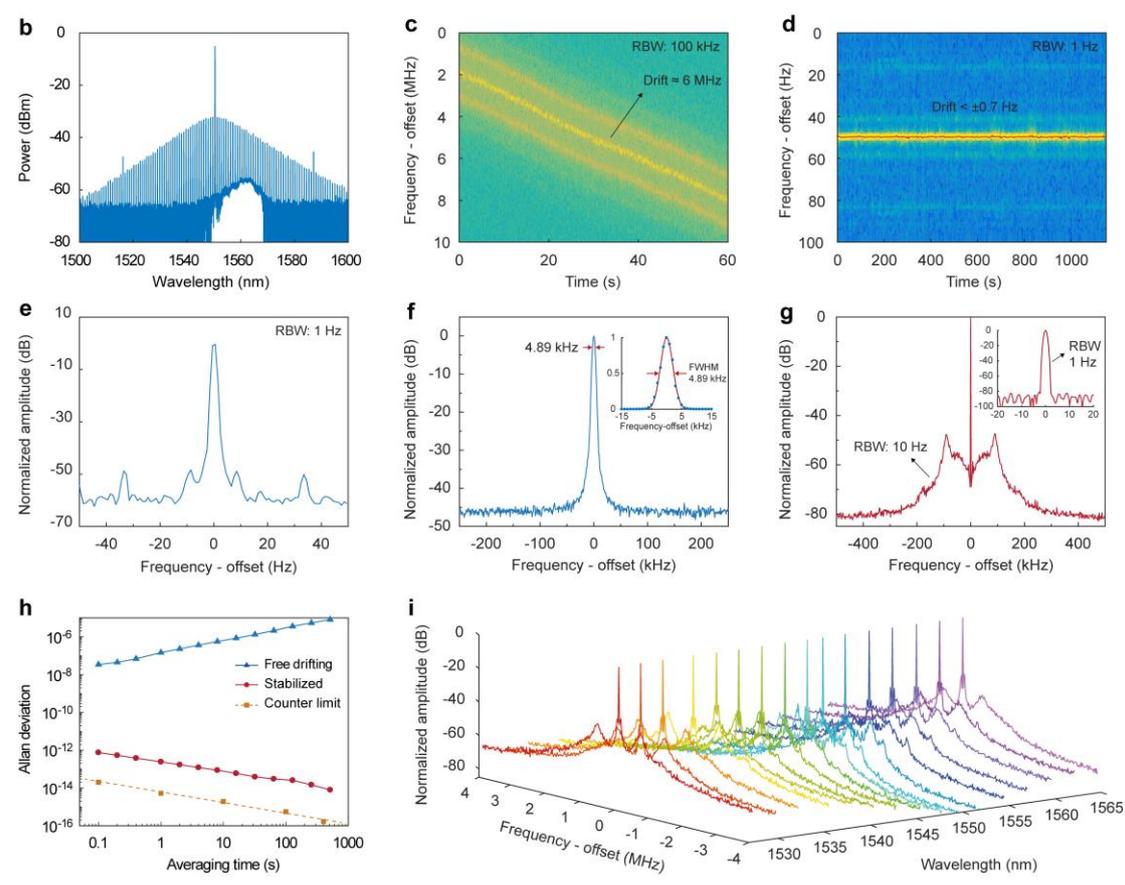

**Fig. 2 Implementation of the frequency-hopping laser source. a,** Experimental setup of the proposed frequency-hopping laser source. A high-power laser enters the microcavity, stimulating the soliton comb, and a filter selects one of the comb lines as a reference for the TLS. Phase-locked loops (loop1 and loop2) ensure the central comb line is locked to the ultra-stable laser and the repetition frequency is locked to the atomic clock, while loop3 locks the TLS to the selected comb line, generating a frequency-hopping laser source. ECDL, external cavity diode laser; EDFA, erbium-doped fiber amplifier; AOM, acousto-optic modulator; FBG, fiber Bragg grating; EOM, electro-optical modulator; OPL, offset phase lock; LPF, low-pass filter; PID, proportional-

integral-derivative; TLS, tunable laser source. **b,** Optical spectrum of the soliton microcomb. **c,** Spectrogram of the beat note between the free-drifting central comb line and the ultra-stable laser. RBW, resolution bandwidth. **d,e,** Spectrogram and spectrum of the beat note between the stabilized central comb line and the ultra-stable laser. **f,** RF spectrum of the free-drift microcomb spacing. **g,** RF spectrum of the stabilized microcomb spacing. **h,** Allan deviation of the free-drifting and the stabilized microcomb spacings. **i,** Spectra of beat notes between the stabilized TLS and comb lines.

To precisely monitor the drift of the comb spacing, the generated microcomb is directed into a Mach-Zehnder modulator (MZM) after its central comb line is partially suppressed by the FBG. The MZM is driven by an RF source with a frequency of ($f_{rep}$ - 10 MHz)/2 that is precisely phase-locked to a rubidium atomic clock. Lauching the modulated signal to a low-speed photodetector with a pre-amplifier, the spectrum information of $f_{rep}$ is transferred to the beat note between the adjacent ±1st-order sidebands. As depicted in Fig. 2f, the RF spectrum of the free-drifting beat note exhibits an FWHM of 4.89 kHz. To stabilize the beat note, an offset phase-locked loop regulates the pump power entering the microresonator via the drive power of the acousto-optic modulator (AOM), resulting in the improved spectral purity of $f_{rep}$, as depicted in Fig. 2g. The FWHM is around 1 Hz, which is mainly limited by the resolution of the electrical spectrum analyzer.

To provide a more comprehensive assessment of the comb repetition stability before and after stabilization, Allan deviation (ADEV) analysis was conducted using an SRS SR620 frequency counter, as illustrated in Fig. 2h. At a 1-second averaging time, the free-drifting comb exhibits relative fluctuations of $1.45 \times 10^{-7}$, while the stabilized one achieves remarkable stability of $2.42 \times 10^{-13}$. The gradual $f_{rep}$ drifts can be attributed to temperature variations, while more rapid fluctuations likely stem from intracavity power variations. The Allan deviation of the counter limit was measured with a coherent signal at 10 MHz and normalized to the 103-GHz mode spacing frequency. It is one order of magnitude lower than the stabilized $f_{rep}$, which ensures the accuracy of $f_{rep}$ measurement.

The fully stabilized soliton microcomb can now serve as an optical reference. Utilizing the third offset phase-locked loop, the wavelength of a TLS can be precisely locked to one comb line of the reference comb, enabling linewidth narrowing and stable frequency hopping. Fig. 2i shows 16 spectral curves of the beat note after the TLS is locked to the comb lines in the wavelength range from 1525 to 1565 nm. Notably, the experiment leveraged a rubidium atomic clock to provide external reference clock signals for all RF sources, ensuring the synchronization of the system.

**Vector spectrometer based on parallel ASG and ASR**
Next, we establish the vector spectrometer (see Fig.3a) based on parallel asymmetric ODSB modulation. Leveraging the proposed frequency-hopping laser, we generate an ultra-narrow linewidth and ultra-stable carrier, which is divided into two portions, serving as carriers for two EO combs. The two EO combs are generated at two phase modulators driven by two RF signals fixed at 20 GHz and 20.003 GHz, respectively. While the first EO comb is modulated by a scanning RF signal at $f_e$, generating five pairs of ±1st-order scanning sidebands, the second EO comb is frequency-shifted by 80 MHz in an AOM. The signals in the two paths are combined and further divided into two branches: one for reference and the other to excite and interrogate the device under test. The above components form a parallel ASG.

The ASR consists of two photodetectors and a series of signal-processing modules. Following square-law detection in the photodetectors, two tunable electrical bandpass filters (BPFs) select the desired beat notes, which are converted into intermediate frequency (IF) signals at mixers and then filtered by BPF2. A high-resolution analog-to-digital converter samples the IF signals, which are then processed in a digital signal processor. RF1, RF2, RF3, and RF4 are synchronized to ensure the coherence of the system. Straightforward calibration, achieved by removing the device under test, eliminates common-mode noise between the measurement and reference paths, leading to a further increase in accuracy.

To verify the ultra-fine frequency scanning capability enabled by the ASG, we selected one frequency-sweeping sideband to beat with an ultra-stable laser. Fig. 3b shows a spectrogram of the beat note obtained by an electrical spectrum analyzer. As can be seen, the sideband performed a sweep in a step of 1 Hz, and the dwell time of each frequency point is about 100 s. This demonstrates the ASG's capability to scan a device under test with a hyperfine frequency step. Combined with the sub-hertz linewidth of the swept sideband, the proposed spectrometer based on ASG and ASR can achieve a Hz-level frequency resolution. To confirm this superiority, we measured the spectrum of a two-tone signal separated by 2 Hz. The two-tone signal is generated by an MZM working at its minimum point, where the modulated carrier is a sub-hertz-linewidth optical signal, and the modulation frequency is 1 Hz. Fig. 3c presents the measured spectrum, indicating that the proposed spectrometer achieves a frequency resolution better than 2 Hz. When the bias of the MZM deviates from the minimum point and the modulation frequency is changed to 5 Hz, the measured spectrum of the MZM output signal is shown in Fig. 3d. As observed, the odd-order sidebands are accurately characterized, with the first-order sidebands located 5-Hz away from the carrier, which is consistent with the modulation frequency. Moreover, the even-order sidebands are sufficiently suppressed, falling below the noise floor, which is consistent with the modulation format.

We also utilize the proposed spectrometer to assess the frequency-domain characteristics of a narrow-linewidth CW laser. A fiber laser (NKT X15) with an intrinsic linewidth of <100 Hz serves as the signal under test. After photodetection, the beat between the signal and the swept sideband generated by the ASG is downconverted, low-pass filtered, and recorded by a digitizer. Compared to the hundred-Hz-level intrinsic linewidth of the signal, the sub-Hz linewidth of the swept sideband can be ignored. Subsequently, utilizing the discrete Fourier transform, the spectral information of the signal under test carried by the downconverted signal is extracted. Fig. 3e shows the measurement of a fiber laser by the proposed spectrometer and a commercial grating spectrometer (Yokogawa AQ6370D) with the highest resolution of 2.5 GHz. As can be seen, the spectral line width measured by the commercial spectrometer is on the order of GHz, failing to reveal the spectral characteristics of the fiber laser with 100-Hz intrinsic linewidth. Our measurement result is much narrower than the commercial spectrometer, which aligns quite well with the fiber laser. The inset shows the power spectral density of the fiber laser's frequency noise, derived from the downconverted signal. The horizontal dashed line marks the white noise level caused by the intrinsic noise. Based on this, the intrinsic linewidth is estimated to be ~87.5 Hz[40], consistent with the laser source datasheet. Fig. 3f shows the fiber laser's frequency drift and ADEV, indicating a signal stability of $2.38\times10^{-10}$ at 1-s averaging time. To verify the vector signal analysis capability of the proposed spectrometer, the signal is sent to a phase modulator driven by a 350-kHz square wave signal. The amplitude and phase (with carrier drift removed) information of the generated phase modulation signal are shown in Fig. 3g. The square-wave signal is well demodulated. To assess the vector device analysis capability, an $H^{13}C^{14}N$ gas cell serves as the device under test. By wavelength scanning of the frequency-hopping laser, a measurement spanning over 33 nm is obtained (Fig. 3h). In this case, the frequency setting resolution is 100 MHz. The measurement results agree very well with the HITRAN[41] simulation and notably outperform that measured by a commercial spectrometer (Yokogawa AQ6370D). Moreover, the higher resolution of the proposed spectrometer enables deeper absorption lines, providing richer details. It is worth noting that in actual measurement, the number of EO comb lines can be adjusted according to the measurement requirements. When a high signal-to-noise ratio and a small measurement range are required, EO combs can be dispensed with by removing the phase modulators (PM1 and PM2).

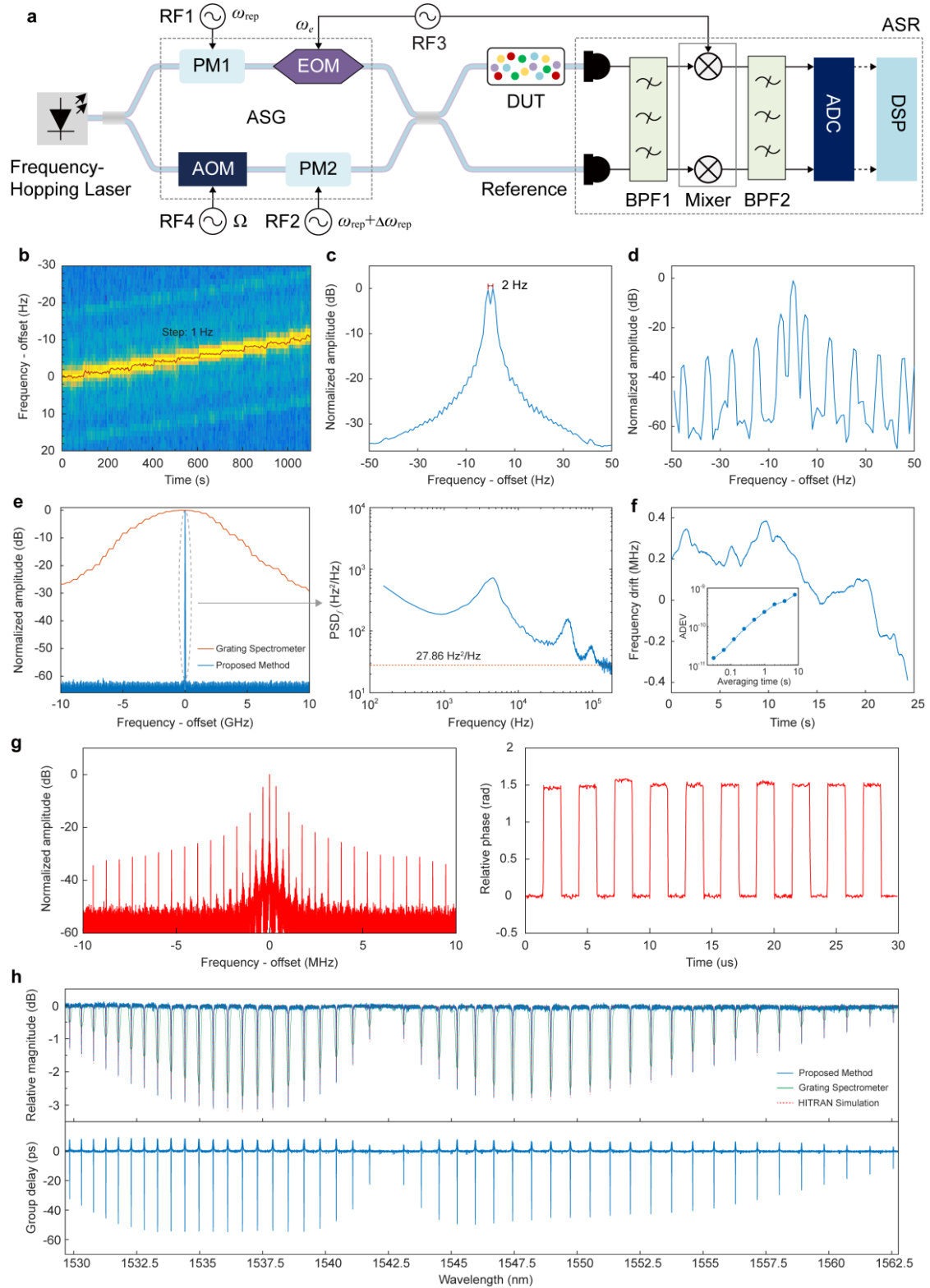

**Fig. 3 Demonstration of vector spectrometer. a,** Experimental setup of the proposed vector spectrometer. The parallel ASG produces two EO combs with slightly different central and repetition frequencies using an AOM and two PMs. One comb undergoes modulation by a frequency-sweeping microwave signal in the EOM via carrier-suppressed modulation, and the other is frequency-shifted in the AOM. The two combs are combined and further divided into measurement and reference branches. The measurement signal carries the information of the device under test, and the ASR precisely extracts the desired information. PM, phase modulator; AOM, acousto-optic modulator; EOM, electro-optical modulator; DUT, device under test; BPF, bandpass filter; ADC, analog-to-digital converter; DSP, digital signal processor. **b,** Spectrogram of the beat note between the sweeping sideband and the ultra-stable laser. **c,** Spectrum of a carrier-suppressed double-sideband modulated signal (modulation signal frequency: 1 Hz) measured by the proposed spectrometer. **d,** Similar to **c**, but for a modulation signal frequency at 5 Hz, with the EOM bias slightly offset from the minimum point. **e,** Measurement of a fiber laser with an intrinsic linewidth of <100 Hz by the proposed spectrometer (blue line) and a commercial grating spectrometer (red line). The inset shows the power spectral density of the laser's frequency noise. The horizontal dashed line marks the white noise level caused by the intrinsic noise, from which the intrinsic linewidth can be estimated to be ~87.5 Hz[40]. **f,** Frequency drift and Allan deviation of the fiber laser source measured by the proposed vector spectrometer. **g,** Measured amplitude spectrum and phase spectrum of a phase-modulated signal using the fiber laser as the carrier. **h,** Measurement results of an $H^{13}C^{14}N$ gas cell in a 33 nm span using the proposed method (blue lines) compared with a commercial optical spectrometer (green line) and HITRAN simulation (red dotted line).

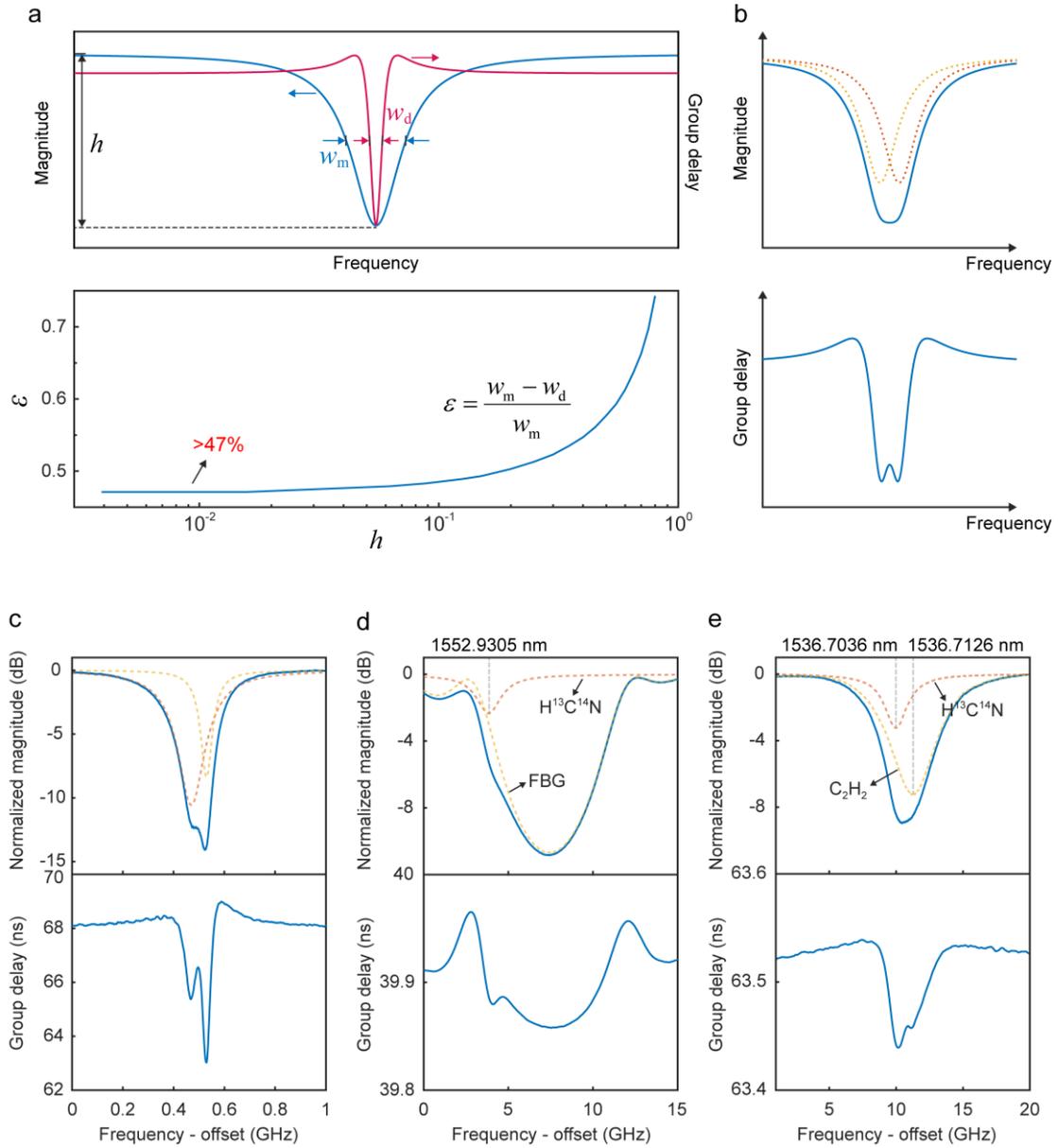

**Fig. 4 Vector spectrometer enables super-recognition. a,** Principle of the super-recognition. **b,** Simulation results depicting the magnitude response and group delay response of a two-component system with two overlapping band-stop responses. **c,** Measured superposition of two adjacent microcavity resonators. **d,** Spectra of $H^{13}C^{14}N$ gas cell and FBG in series measured by the proposed spectrometer. **e,** Measured spectra resulting from superimposing the $H^{13}C^{14}N$ gas cell response and $C_2H_2$ gas cell response. Dashed lines represent the spectral responses of the two components measured separately, while solid lines represent the spectral responses measured together.

## Super-recognition of matters

Interactions between light and matter are fundamental in spectroscopy, sensing, and laser technologies. Traditionally, measuring intensity changes in light-matter interaction is the primary method for the identification of matter. For example, spectroscopy uses absorbance with respect to wavelength for qualitative analysis and quantification. However, relying solely on intensity information can pose challenges in spectrometric analysis, particularly in scenarios with overlapping absorption spectra in a multi-component system. The FWHM of the absorption line significantly influences the separation of overlapping spectral lines: the narrower the FWHM, the easier it becomes to distinguish between two adjacent absorption lines. Here, we present an original method that uses group delay variations during light-matter interactions for spectral analysis. As can be seen from Fig. 4a, the group delay spectrum corresponding to the Lorentz-shaped absorption line exhibits a significantly narrower FWHM. In addition, as the absorption line deepens, the group delay spectral line narrows by at least 47%. Fig. 4b shows the simulation results of the magnitude response and group delay response of a two-component system. Obviously, while the two magnitude lines overlap extensively, the group delay response can still show two distinct peaks. Thus, compared to magnitude responses, group delay responses offer significant advantages in separating overlapping spectral lines in a multi-component system.

To experimentally demonstrate the resolving capability of the group delay response, we measured overlapping spectral lines using the proposed vector spectrometer. It is worth noting that accurate measurement of the group delay response, derived from phase with respect to frequency, demands a high SNR to achieve precise phase, a high spectral resolution to maintain frequency precision, and an acceptable measurement time to ensure reliable results. The experimental verification of the effect of spectral resolution and SNR on the group delay response is presented in the Supplementary Materials. The proposed vector spectrometer meets these measurement requirements effectively. In the first validation experiment, a microresonator, an FBG, an $H^{13}C^{14}N$ gas cell (20 Torr, 16.5 cm), and a $C_2H_2$ gas cell (40 Torr, 6.0 cm) are used as the devices under test. Fig. 4c illustrates the measured spectral response of two adjacent resonators of the microresonator. While the two valleys in the magnitude spectrum are barely distinguishable, the two spectral lines can be clearly separated in the group delay response. Fig. 4d shows the measurement of the overlapped spectra near 1552.9 nm introduced by the FBG and $H^{13}C^{14}N$ gas cell. When the absorption line of $H^{13}C^{14}N$ is located at the edge of the notch of the FBG, discerning them in the magnitude response proves challenging despite the high spectral resolution. Conversely, in the group delay response, we can distinctly observe the two notches. Near 1536.7 nm, where the absorption line of the $H^{13}C^{14}N$ becomes closer to that of $C_2H_2$, the measured magnitude response is completely overlapped and cannot be directly distinguished, as depicted in Fig. 4e. In contrast, we can clearly discern the two notches in the group delay response.

The group delay response represents a fundamental physical parameter that can be measured in a variety of ways, and the device under test can be anything from gas molecules to optical chips. This implies that the new measurement paradigm that uses group delay response to distinguish overlapping spectral lines holds significant potential for broad-scale applications. In comparison to indirect measurement techniques such as multidimensional coherent spectroscopy[42], which relies on external perturbations to stimulate varied responses from resonances, the group delay response offers advantages of simplicity, directness, and real-time analysis. Drawing an analogy from psychology, where a super-recognizer denotes an individual with exceptional face recognition abilities[43], we use "super-recognition" to highlight the excellent ability of group delay response in discerning and identifying overlapping spectral lines originating from diverse sources.

**Conclusion**

The mutual constraint of spectral resolution, wavelength range, and accuracy poses a significant challenge in spectral measurement. To address this, we have introduced an ultrastable microcomb as an absolute frequency reference to establish a rapid frequency-hopping laser with a broad wavelength range and an exceptionally narrow linewidth. The frequency-hopping laser's stability with deviations of less than ±0.7 Hz over 1000 s, linewidth of approximately 1 Hz, and spectral manipulation finesse of 1 Hz enables spectral measurements with a resolution of up to 2 Hz. This represents a four-order-of-magnitude enhancement compared to state-of-the-art techniques and a six-order-of-magnitude improvement over commercial spectrometers. We have also proposed a vector spectral analysis method based on parallel ASG and ASR, allowing for ultra-high-resolution frequency sweeps while minimizing measurement errors and ensuring high precision. By integrating wideband frequency hopping with parallel ASG and ASR, we achieve broadband measurements beyond 33 nm. These findings underscore the unprecedented functionality of vector measurement. Furthermore, we unveil the remarkable "super-recognition" potential of group delay response in discerning overlapping spectral lines, providing a more efficient tool for identifying different matters than traditional magnitude-based approaches.

**Future perspectives**

Our upcoming challenge involves integrating the discussed technique into miniature integrated spectrometers[44-50], with the aim of achieving optimal performance and cost-effectiveness. This entails the integration of both the stable microcomb and parallel ASG/ASR. The resolution and wavelength range can be customized to align with specific application scenarios, with the potential to extend the range to encompass other spectral bands like visible or infrared. The super-recognition capabilities of the group delay response are poised to inaugurate a new era in vector measurement. We foresee that our work will catalyze a multitude of possibilities in the field of vector measurement. For example, leveraging the group delay response for distance measurement or exploring the integration of a spectrometer with LIDAR systems could unlock innovative capabilities and drive exciting advancements across various domains.

**Methods**

**Implementation of the Frequency-hopping laser**

The experimental setup of the soliton microcomb generation and stabilization is shown in Fig. 2a. The soliton microcomb was generated by pumping the $Si_3N_4$ microresonator fabricated by Ligentec with an external cavity diode laser (Toptica CTL 1550). The pump was amplified by a high-power EDFA and was manually tuned from the blue detuning side into the resonance. Using two offset PLLs, the central comb line and the repetition frequency of the microcomb were locked to an ultra-stable laser (MenloSystems ORS) and an atomic clock (SRS PRS10), respectively. A PIN photodiode (Discovery Semiconductors DSC50S) generated the beat note of the central comb line and the ultra-stable laser, which was then amplified by a low-noise RF amplifier (Mini-Circuits ZX60-33LNR+). The beat note was transmitted to the offset PLL to lock the central comb line. The PLL consisted of a double-balanced mixer (Mini-Circuits ZX05-5-S+), a low-pass filter (Mini-Circuits SLP-21.4+), and a high-speed PID controller (Toptica FALC 110). A microwave source (R&S SMAB-B131) provided the RF reference to the offset PLL, controlling the frequency spacing between the pump and the ultra-stable laser. The spectrogram of the beat note between the central comb line and the ultra-stable laser was measured by a phase noise analyzer (R&S FSWP50). A high-speed intensity modulator (EOSPACE AX-0MVS-65-LV) with a 6-dB bandwidth of 65 GHz and a low-speed photodetector (Koheron PD100) with a 3-dB bandwidth of approximately 100 MHz constituted a microwave photonic downconverter for converting the mode spacing into a sinusoidal signal with a frequency of approximately 10 MHz. The intensity modulator was driven by an RF signal with a frequency from a 67-GHz microwave source (R&S SMAB-B167). Since the down-conversion signal from the low-speed photodetector is relatively weak, we used a 40-dB gain amplifier and a bandpass filter with a bandwidth of 1 MHz to amplify and filter it. The amplified signal was then split into two parts. One was sent to another offset PLL with a similar configuration, and the other was monitored by the FSWP50 and a frequency counter (SRS SR620). The second PLL employed a low-frequency signal from an RF generator (Rigol DG4202) as an external reference and produced a feedback signal. The feedback signal was led to the amplitude modulation input port of another microwave source (Agilent E4421B), to control the power of its 80-MHz output signal. The 80-MHz signal drives an AOM to achieve stable comb spacing by controlling the pump power. Finally, following the same procedure as stabilizing the central comb line, we lock another CTL 1550 to any desired comb mode with a frequency difference of $f_{ofs}$. By tuning the CTL 1550 output frequency and fine-tuning the PLL reference frequency, the frequency spacing $f_{ofs}$ can be adjusted, allowing the desired optical frequency $f_o = f_{cent} + N \cdot f_{rep} + f_{ofs}$ to be obtained.

**Measurement of optical signals**

Figure 3a shows the experimental setup of the measurement system. The parallel ASG comprised two phase modulators (PM, EOSPACE), a Mach-Zehnder modulator (Fujitsu FTM7938EZ), and an acousto-optic modulator (G&H PM FIBER-Q). The two phase modulators were driven by high-power RF signals to produce two EO combs with repetitions of 20 GHz and 20.003 GHz, respectively. The Mach-Zehnder modulator was biased at the minimum point, while the acousto-optic modulator served as an 80 MHz frequency shifter. The parallel ASR consisted of two PIN photodiodes (Finisar 2120RA) and an electrical receiver. The electrical receiver comprised a series of RF filters, mixers, analog-to-digital converters, and signal processing modules. One can also directly use instruments such as spectrum analyzers (R&S FSWP50), frequency counters (SRS SR620), and vector network analyzers (R&S ZVA67) to act as the receiver. Extended Data Figure 2 shows the schemes for generating the optical signals under test shown in Fig. 3. In Extended Data Fig. 1a, the frequency scanning resolution of the ASG was verified by beating with an ultra-stable laser. The spectrogram of the beat note was recorded by FSWP50. We also verified the frequency resolution of the vector spectrometer by modulating the ultra-stable laser to produce multi-tone signals with ultra-narrow linewidth and close frequencies, as shown in Extended Data Fig. 1b. An arbitrary waveform generator (AWG) outputs a low-frequency sinusoidal signal with DC bias to the DC port of the electro-optical modulator (EOM, EOSPACE) to achieve low-frequency modulation and bias point setting. By adjusting the bias voltage of the AWG output, the EOM is biased at the minimum point. Then, we set the AWG output frequency to 1 Hz, and obtained the two-tone signal shown in Fig. 3c. Also, changing the AWG output bias and setting the output frequency to 5 Hz yielded the multi-tone signal shown in Fig. 3d. Furthermore, to demonstrate the analytical capability of the proposed vector spectrometer for the amplitude and phase of a free-drifting laser, we performed measurements using the setups as shown in Extended Data Figs. 1c and 1d. A fiber laser (NKT X10) with an intrinsic linewidth of <100 Hz served as the laser under test, and a phase modulator (PM, EOSPACE), driven by a square wave from the AWG, was used to generate the optical phase modulation signal under test. Extended Data Fig. 2 illustrates the original demodulated phase and the phase drift. Through phase drift compensation and noise reduction, the data shown in Fig. 3g was obtained.

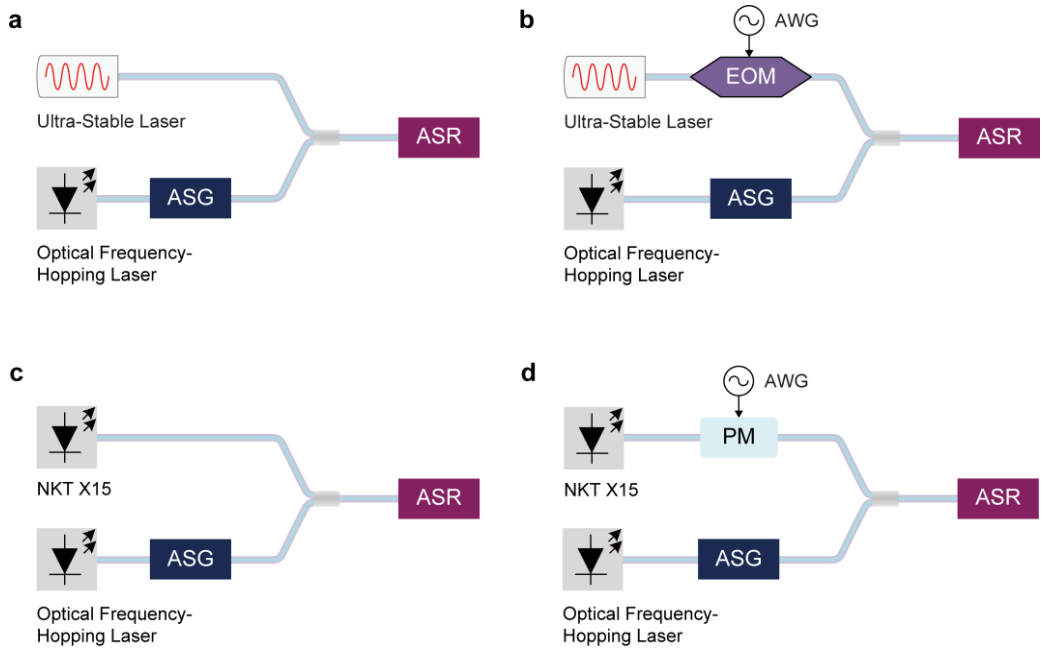

Extended Data Fig. 1 | Experimental setups for generating the optical signals under test.

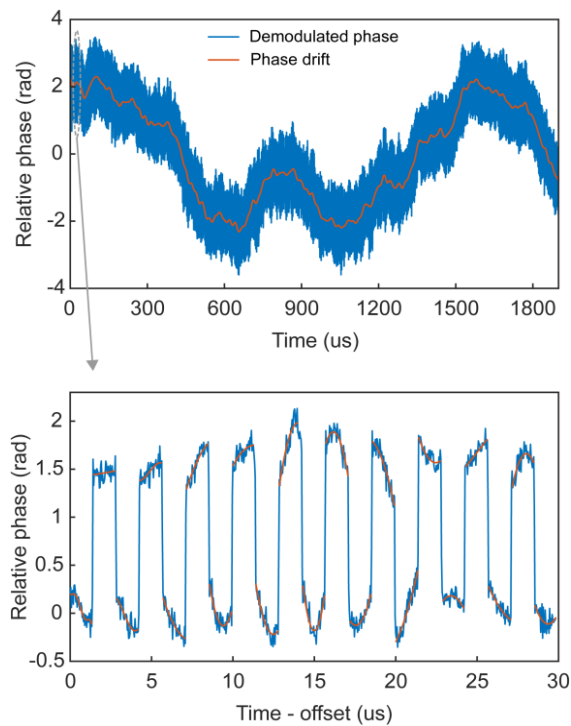

Extended Data Fig. 2 | Demodulated phase of an optical phase modulation signal.

**The impact of frequency resolution on group delay measurement**

High frequency scanning resolution helps to improve the precision of group delay measurement. This arises from the ability to achieve a large amount of phase response data in the linear region, facilitating a least squares estimate of the group delay. Mathematically, the standard deviation of the group delay calculated from the phase difference is represented as $\sigma_{\text{diff}} = \text{sqrt}(2) \cdot \sigma / \Delta\omega$, where $\sigma$ denotes the mean square error of the phase response, and $\Delta\omega$ is the frequency spacing. With the ultra-high resolution offered by the proposed vector spectrometer, we can perform a finer scan in the frequency span to obtain *n*-point phase response. The standard deviation of the estimated group delay can be expressed as

$$\sigma_{LSE} = \frac{\sigma}{\left(\frac{\Delta\omega}{n-1}\right)\sqrt{\sum_{i=1}^{n}\left((i-1)-\overline{(i-1)}\right)^2}} \tag{1}$$

We define the ratio of $\sigma_{\text{diff}}$ and $\sigma_{\text{LSE}}$ as the precision improvement factor, which can be calculated as $Q = \sigma_{\text{diff}}/\sigma_{\text{LSE}} = \text{sqrt}(n(n+1)/(n-1)/6)$. It illustrates that the precision of the group delay is enhanced with increased resolution. For example, when the resolution improves by 100 times, the accuracy of the group delay is enhanced by 4.2 times, and with a resolution improvement of 1000 times, the enhancement could rise to 12.9 times.

**Acknowledgments**

This work was supported in part by the National Natural Science Foundation of China (62271249).

**Author contributions**

T.Q. and S.L.P. initiated the project. T.Q., S.P.L., and S.L.P. conceived the research and method. S.P.L. and T.Q. designed and set up the experiments. Under the supervision of S.L.P. and Y.G.L., S.P.L. and H.S.Y. performed optical frequency-hopping laser generation; S.P.L. carried out the active device measurements; L.H.W., Y.J.F., X.H.T., S.P.L., T.Q., and M.H.C. carried out the passive device measurements. J.J.H. and J.Q.L. fabricated the microcavity. S.P.L. and T.Q. analyzed and interpreted the data. T.Q., S.P.L., and S.L.P. wrote the manuscript. J.M.L and Y.G.L. provided intellectual inputs and edited the manuscript.

**Data availability**

The data supporting the findings of this study can be obtained from the corresponding author.

**References**

[1] Preussler, S., & Schneider, T. Attometer resolution spectral analysis based on polarization pulling assisted Brillouin scattering merged with heterodyne detection. *Opt. Express* **23**, 26879-26887 (2015).
[2] Yang, Q. F. et al. Vernier spectrometer using counterpropagating soliton microcombs. *Science* **36**, 965-968 (2019).
[3] Lemarchand, C. et al. Progress towards an accurate determination of the Boltzmann constant by Doppler spectroscopy. *New J. Phys.* **13**, 073028 (2011).
[4] Schulz, H., & Baranska, M. Identification and quantification of valuable plant substances by IR and Raman spectroscopy. *Vib. Spectrosc.* **43**, 13-25 (2007).
[5] Vollmer, F., & Yang, L. Review Label-free detection with high-Q microcavities: a review of biosensing mechanisms for integrated devices. *Nanophotonics* **1**, 267-291 (2012).

[6] Qing, T., Li, S., Tang, Z., Gao, B., & Pan, S. Optical vector analysis with attometer resolution, 90-dB dynamic range and THz bandwidth. *Nat. Commun.* **10**, 5135 (2019).

[7] Savchenkov, A. A., Matsko, A. B., Ilchenko, V. S., & Maleki L. Optical resonators with ten million finesse. *Opt. Express* **15**, 6768-6773 (2007).

[8] Bigelow, M. S., Lepeshkin, N. N., & Boyd, R. W. Observation of ultraslow light propagation in a ruby crystal at room temperature. *Phys. Rev. Lett.* **9**, 113903 (2003).

[9] Kwon, D., Jeon, I., Lee, W. K., Heo, M. S., & Kim, J. Generation of multiple ultrastable optical frequency combs from an all-fiber photonic platform. *Sci. Adv.* **6**, eaax4457 (2020).

[10] Jin, W. et al. Hertz-linewidth semiconductor lasers using CMOS-ready ultra-high-Q microresonators. *Nat. Photon.* **15**, 346-353 (2021).

[11] Kikuchi, K. Fundamentals of coherent optical fiber communications. *J. Lightwave Technol.* **34**, 157–179 (2015).

[12] Armani, A. M. et al. Label-free, single-molecule detection with optical microcavities. *Science* **317**, 783-787 (2007).

[13] Ludlow, A. D., Boyd, M. M., Ye, J., Peik, E. & Schmidt, P. O. Optical atomic clocks. *Rev. Mod. Phys.* **87**, 637–701 (2015).

[14] Suh, M. G., & Vahala, K. J. Soliton microcomb range measurement. *Science* **359**, 884-887 (2018).

[15] Liu, J. et al. Photonic microwave generation in the X- and K-band using integrated soliton microcombs. *Nat. Photon.* **14**, 486–491 (2020).

[16] Kibble, B. P., Rowley, W. R. C., Shawyer, R. E., & Series, G. W. An experimental determination of the Rydberg constant. *J. Phys. B: Atom. Mol. Phys.* **6**, 1079 (1973).

[17] Kneubühl F. Diffraction Grating Spectroscopy. *Appl. Opt.* **8**, 505-519 (1969).

[18] Palmer, C.& Loewen, E. G. *Diffraction Grating Handbook* (Newport Corporation, New York, 2005).

[19] Agilent Technologies. *Optical Spectrum Analysis* (1996).

[20] Griffiths, P. R. Fourier Transform Infrared Spectrometry. *Science* **222**, 297-302 (1983).

[21] Kita, D.M. et al. High-performance and scalable on-chip digital Fourier transform spectroscopy. *Nat. Commun.* **9**, 4405 (2018).

[22] Hernandez, G. J. *Fabry-Perot Interferometers* (Cambridge University Press, Cambridge, 1988).

[23] Wang, X., Chen, C., Pan, L. & Wang, J. A graphene-based Fabry-Pérot spectrometer in mid-infrared region. *Sci. Rep.* **6**, 32616 (2016).

[24] APEX Technologies. *Ultra High Resolution OSA/OCSA for Characterizing and Evaluating Optical Frequency Comb Sources* (2013).

[25] Preußler, S., Wiatrek, A., Jamshidi, K. & Schneider, T. Brillouin scattering gain bandwidth reduction down to 3.4 MHz. *Opt. Express* **19**, 8565–8570 (2011).

[26] Xue, M., Zhao, Y., Gu, X. & Pan, S. Performance analysis of optical vector analyzer based on optical single-sideband modulation. *J. Opt. Soc. Am. B* **30**, 928–933 (2013).

[27] Gifford, D. K., Soller, B. J., Wolfe, M. S. & Froggatt, M. E. Optical vector network

analyzer for single-scan measurements of loss, group delay, and polarization mode dispersion. *Appl. Opt.* **44**, 7282 (2005).

[28] Bao, Y. et al. A digitally generated ultrafine optical frequency comb for spectral measurements with 0.01-pm resolution and 0.7-μs response time. *Light Sci. Appl.* **4**, e300 (2015).

[29] Román, J. E., Frankel, M. Y., & Esman, R. D. Spectral characterization of fiber gratings with high resolution. *Opt. Lett.* **23**, 939–941 (1998).

[30] Voges, E., Ostwald, O., Schiek, B. & Neyer, A. Optical phase and magnitude measurement by single sideband homodyne detection. *IEEE J. Quantum Electron.* **18**, 124–129 (1982).

[31] Morozov, O. et al. Ultrahigh-resolution optical vector analyzers. *Photonics* **7**, 14 (2020).

[32] Li, W., Wang, W., Wang, L., & Zhu, N. Optical vector network analyzer based on single-sideband modulation and segmental measurement. *IEEE Photon. J.* **6**, 1–8 (2014).

[33] Qing, T., Xue, M., Huang, M. & Pan, S. Measurement of optical magnitude response based on double-sideband modulation. *Opt. Lett.* **39**, 6174 (2014).

[34] Qing, T. et al. Comprehensive vector analysis for electro-optical, opto-electronic, and optical devices. *Opt. Lett.* **46**, 1856-1859 (2021).

[35] Udem, T., Holzwarth, R. & Hänsch, T. W. Optical frequency metrology. *Nature* **416**, 233–237 (2002).

[36] Cundiff, S. T. & Ye, J. Colloquium: femtosecond optical frequency combs. *Rev. Mod. Phys.* **75**, 325–342 (2003).

[37] Kippenberg, T. J., Gaeta, A. L., Lipson, M. & Gorodetsky, M. L. Dissipative Kerr solitons in optical microresonators. *Science* **361**, eaan8083 (2018).

[38] Marin-Palomo, P. et al. Microresonator-based solitons for massively parallel coherent optical communications. *Nature* **546**, 274–279 (2017).

[39] Liu, J. et al. Monolithic piezoelectric control of soliton microcombs. *Nature* **583**, 385–390 (2020).

[40] Di Domenico, G., Schilt, S., & Thomann, P. Simple approach to the relation between laser frequency noise and laser line shape. *Appl. Opt.*, **49**, 4801-4807 (2010).

[41] Gordon, I. E. et al. The HITRAN2016 molecular spectroscopic database. *J. Quant. Spectrosc. Radiat. Transf.* **203**, 3-69 (2017).

[42] Lomsadze, B. & Cundiff, S. T. Frequency combs enable rapid and high-resolution multidimensional coherent spectroscopy. *Science* **357**, 1389-1391 (2017).

[43] Nador, J. D., Zoia, M., Pachai, M. V. & Ramon, M. Psychophysical profiles in super-recognizers. *Sci. Rep.* **11**, 13184 (2021).

[44] Bacon, C. P., Mattley, Y. & DeFrece, R. Miniature spectroscopic instrumentation: applications to biology and chemistry. *Rev. Sci. Instrum.* **75**, 1–16 (2004).

[45] Crocombe, R. A. Portable spectroscopy. *Appl. Spectrosc.* **72**, 1701–1751 (2018).

[46] McGonigle, A. J. S. et al. Smartphone spectrometers. *Sensors* **18**, 223 (2018).

[47] Yang, Z., Albrow-Owen, T., Cai, W. & Hasan, T. Miniaturization of optical spectrometers. *Science* **371**, eabe0722 (2021).

[48] Yao, C. et al. Integrated reconstructive spectrometer with programmable photonic circuits. *Nat. Commun.* **14**, 6376 (2023).

[49] Li, A. et al. Advances in cost-effective integrated spectrometers. *Light Sci. Appl.* **11**, 174 (2022).

[50] Redding, B., Liew, S., Sarma, R. & Cao, H. Compact spectrometer based on a disordered photonic chip. *Nat. Photon.* **7**, 746–751 (2013).